\documentclass[12pt]{article}

\usepackage{fullpage}

\usepackage[english]{babel}

\RequirePackage{amsthm,amsmath,amsfonts,amssymb}
\RequirePackage[authoryear]{natbib}
\usepackage{extarrows}

\usepackage{adjustbox}
\usepackage{algorithm}
\usepackage{caption}
\usepackage{algpseudocode}

\usepackage{ulem}
\usepackage{array}

\usepackage[english]{babel}
\usepackage{bigints}
\usepackage{booktabs}

\usepackage{enumerate}
\usepackage{epstopdf}
\usepackage{graphicx,psfrag,epsf}
\usepackage{float}
\usepackage{afterpage}

\usepackage{listings}
\usepackage{lscape}
\usepackage{multirow}
\usepackage{makecell}

\usepackage{rotating}
\usepackage[rightcaption]{sidecap}
\usepackage[flushleft]{threeparttable}
\usepackage{wrapfig}
\usepackage[table,xcdraw]{xcolor}
\usepackage{soul}

\usepackage[hidelinks,colorlinks=true,linkcolor=blue,urlcolor=blue, citecolor=blue,anchorcolor=blue]{hyperref}

\newtheorem{theorem}{Theorem}

\newtheorem{lemma}[theorem]{Lemma}

\newtheorem{innercustomgeneric}{\customgenericname}
\providecommand{\customgenericname}{}
\newcommand{\newcustomtheorem}[2]{%
  \newenvironment{#1}[1]
  {%
   \renewcommand\customgenericname{#2}%
   \renewcommand\theinnercustomgeneric{##1}%
   \innercustomgeneric
  }
  {\endinnercustomgeneric}
}

\newcustomtheorem{customthm}{Theorem}
\newcustomtheorem{customlemma}{Lemma}

\definecolor{yuancolor2}{rgb}{0.0, 0.42, 0.24}
\newcommand{\yuan}[1]{{#1}}
\definecolor{crimsonglory}{rgb}{0.75, 0.0, 0.2}
\definecolor{darkorchid}{rgb}{0.6, 0.2, 0.8}

\providecommand{\keywords}[1]
{
  \small	
  \textbf{\textit{Keywords:}} #1
}

\title{\Large 
Evaluating the effects of high-throughput structural neuroimaging predictors on whole-brain functional connectome outcomes via network-based vector-on-matrix regression }

\date{}
\author{\small Tong Lu$^{1}$, Yuan Zhang$^{2}$, Vince Lyzinski$^{1}$, Chuan Bi$^{3}$, Peter Kochunov$^{4}$, Elliot Hong$^{4}$, Shuo Chen$^{3,5}$\\
\footnotesize
$^{1}$Department of Mathematics, University of Maryland,
College Park\\
\footnotesize
$^{2}$ Department of Statistics, The Ohio State University\\
\footnotesize
$^{3}$Maryland Psychiatric Research Center, School of Medicine, University of Maryland \\
\footnotesize
$^{4}$Department of Psychiatry and Behavioral Science,  University of Texas Health Science Center\\
\footnotesize
$^{5}$Division of Biostatistics and Bioinformatics, School of Medicine, University of Maryland \\
\footnotesize *\href{mailto:
chen@som.umaryland.edu}{shuochen@som.umaryland.edu}
}
\vspace{-3mm}

\begin{document}
\maketitle
\begin{sloppypar}

\begin{abstract}
The joint analysis of multimodal neuroimaging data is critical in the field of brain research because it reveals complex interactive relationships between neurobiological structures and functions. In this study, we focus on investigating the effects of structural imaging (SI) features, including white matter micro-structure integrity (WMMI) and cortical thickness, on the whole brain functional connectome (FC) network. To achieve this goal, we propose a network-based vector-on-matrix regression model to characterize the FC-SI association patterns. We have developed a novel multi-level dense bipartite and clique subgraph extraction method to identify which subsets of spatially specific SI features intensively influence organized FC sub-networks. The proposed method can simultaneously identify highly correlated structural-connectomic association patterns and suppress false positive findings while handling millions of potential interactions.  We apply our method to a multimodal neuroimaging dataset of 4,242 participants from the UK Biobank to evaluate the effects of whole-brain WMMI and cortical thickness on the resting-state FC.  The results reveal that the WMMI on corticospinal tracts and inferior cerebellar peduncle significantly affect functional connections of sensorimotor, salience, and executive sub-networks with an average correlation of 0.81  ($p<0.001$).   \end{abstract}




\section{Introduction}
\label{MOAT_Intro}

Neuroimaging data play a fundamental role in deciphering the operations of the human brain, the most complex organ. These data come in various modalities, including magnetic resonance imaging (MRI), diffusion tensor imaging (DTI), and functional MRI (fMRI). 
Each modality reveals distinct aspects of the brain's structure and functionality. 
For example, MRI provides high-resolution images of the brain's structure, offering valuable physical information such as size, shape, and cortical thickness. 
DTI assesses the integrity of white matter microstructures by calculating fractional anisotropy. 
The fMRI data capture dynamic blood flow changes in different brain regions to measure localized neural activity and functional connections.

In statistical analysis, neuroimaging data are commonly represented in two forms: \emph{vectors} (e.g., a list of region-wise cortical thickness measures) and \emph{association matrices} (e.g., functional connectivity strengths stored in a weighted adjacency matrix) \citep{bullmore2009complex, wig2014approach, fornito2016fundamentals,wang2023variational}.
Instead of studying brain structural imaging (SI) and functional connectivity (FC) data separately, exploring their intricate interplay could significantly deepen our understanding of the brain, including its development and aging \citep{smith2004advances, drevets2008brain, bowman2012determining, kemmer2018evaluating}. 
For example, brain regions connected by white matter tracts with higher fractional anisotropy are more likely to demonstrate strong FCs, which, in turn, can influence cognitive processes such as attention, memory, and decision-making.

There exists little work on the joint analysis of multi-modal neuroimaging data despite its clear importance, possibly due to the challenge presented by ultra-high dimensionality and intertwined data structures.
In conventional brain connectome studies, researchers frequently collect $10^5$ FC measures across hundreds of brain regions and up to $10^4$ SI measures, resulting in billions ($10^9$) of FC-SI pairs. 
This not only creates significant computational demands but also poses challenges for multiple-testing correction.
Traditional correction methods like the false discovery rate (FDR) and family-wise error rate (FWER) often yield almost no supra-threshold FC-SI pairs, as demonstrated by extensive simulation studies.
Moreover, FC and SI display data structures indicative of certain connectomic network space and spatial dependence, respectively.
A joint FC-SI analysis needs to incorporate these intertwined data structures into comprehensive statistical modeling, thus producing biologically plausible and interpretable results.
Specifically, our goal in this work is to identify an array of SI variables that intrinsically influences a group of FCs within a brain connectome sub-network, rather than those randomly distributed across the whole-brain connectome, referred to as a systematic pattern of associations.
These challenges underscore the necessity of developing a joint analysis method to address the complexity of multi-modal neuroimaging data.

Recently, advanced statistical methods have been developed to jointly model two sets of neuroimaging features by leveraging techniques including regularization, low rank, and projection models \citep{wang2011sparse, li2012sparse, zhu2014bayesian, kong2019l2rm}.
Many of these methods have been successfully applied to multi-modal imaging data analysis and yielded interesting findings \citep{hayden2006patterns, ball2017multimodal, wehrle2020multimodal, zhang2022identification}. These statistical methods can be broadly classified into two categories. The first category uses regularization-based methods \citep{zhou2014regularized, zhu2017novel, wang2020critic}, where a major limitation of these methods is that the sparsely selected associations fail to take into account the systematic network-level impacts of SIs on FC networks. The second category employs dimensional reduction strategies, such as principal component analysis (PCA) \citep{hotelling1933analysis, jolliffe2016principal, chachlakis2019l1}, which first projects both FCs and SIs into a handful of top principal components and then performs regression analysis on these selected components. However, as an unsupervised dimension reduction technique, PCA-based analysis often extracts fewer associated principal components of outcomes and predictors, thereby missing the truly associated FC-SI pairs.
Sparse canonical correlation analysis (sCCA) methods can be considered as an integration of these two categories and have been widely used in neuroimaging studies \citep{witten2009penalized,lin2013group, uurtio2019large}.  Yet, sCCA methods usually focus on vector-to-vector association analysis, which may also overlook the systematic vector-to-network association patterns that are of particular interest in this work (i.e., the associations between the SI vector and FC sub-connectome represented as a matrix). 
To bridge the methodological gap in modeling vector-to-matrix associations and incorporating latent network structures, we propose a new \textbf{m}ulti-level netw\textbf{o}rk associ\textbf{a}tion me\textbf{t}hod (\textbf{MOAT}) to systematically investigate the FC-SI association patterns.

\begin{figure}[h!]
    \centering
    \includegraphics[width=\textwidth]{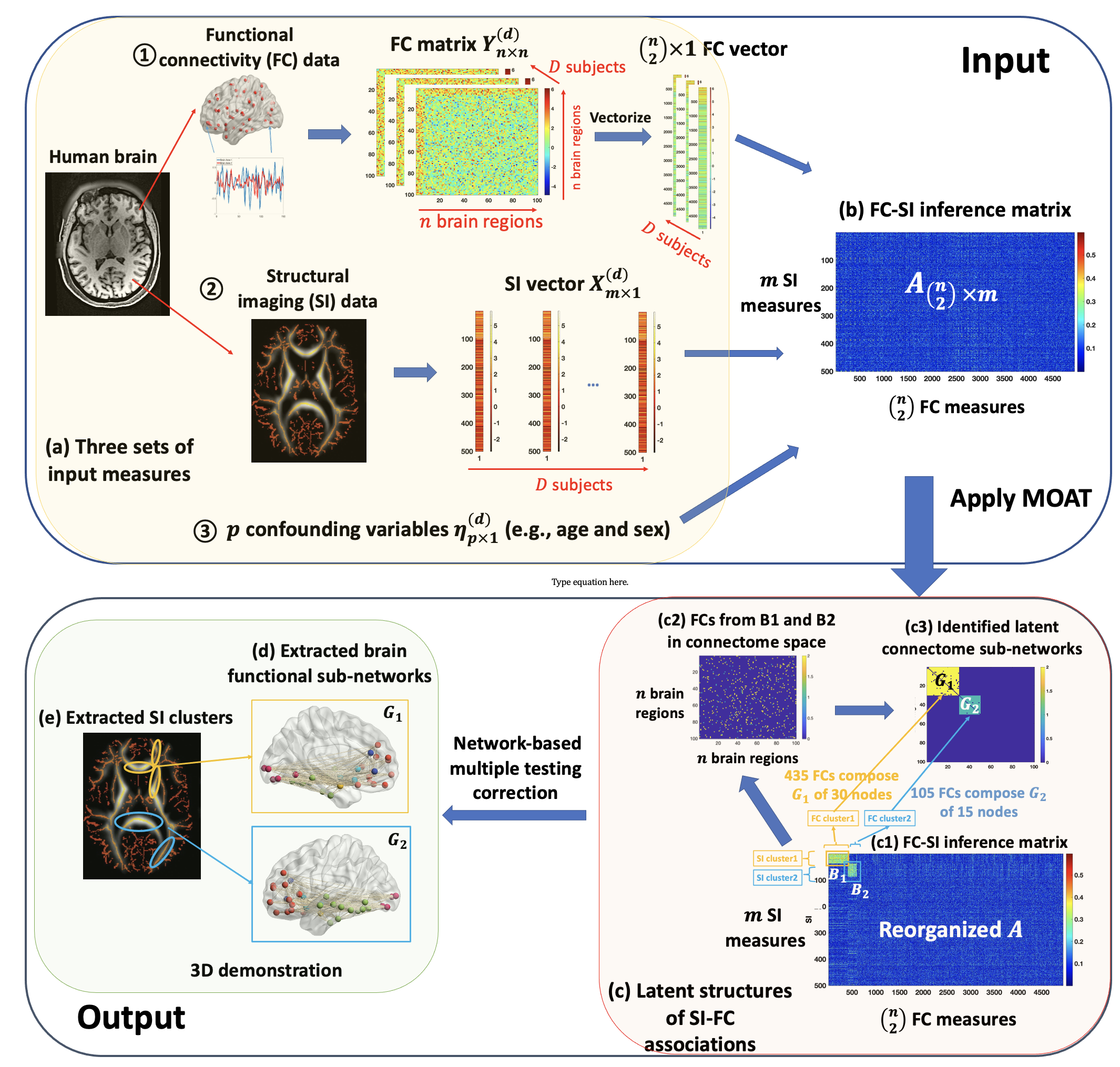}
\caption[The detection pipeline of systematic FC-SI association patterns by MOAT]{
\footnotesize \textit{The detection pipeline of systematic FC-SI association patterns by MOAT. (a) shows three sets of input measures: FC data, SI data, and confounding variables. (b) shows a heatmap of an association matrix $\boldsymbol{A}$, with each element quantifying the association between SI and FC measures (e.g., correlation, test statistics, $-log(p)$ value). A hotter point indicates a stronger FC-SI association. (c) shows the systematic FC-SI association patterns identified by MOAT: the first level (c1) depicts the revealed association patterns between SIs and FCs, where strong associations are mostly concentrated in two dense sub-networks; (c2) shows the FCs in brain connectome space from the detected clusters $B_1$ and $B_2$ in (c1); (c3) shows the second level, which depicts the organized structure of a graph formed by the selected FC clusters in (c2).
Next, our proposed network-based inference test is performed on each FC-SI sub-network identified by MOAT. (d) and (e) shows the 3D visualization of the significant FC-SI network-level associations that pass the inference test.}}
\label{tab:MOAT_pipeline}
\end{figure}

{\autoref{tab:MOAT_pipeline} presents an overview of the MOAT method, which is constructed based on a multi-level graph for structural-functional neuroimaging data. The first level is a bipartite graph that depicts the association patterns between the SI vector as predictors and vectorized FC outcomes, adjusted for other confounding covariates (\autoref{tab:MOAT_pipeline}(a)). Meanwhile, the second level is a complete unipartite graph that reconstructs the vectorized FCs back to a whole-brain connectome network. This multi-level structure enables the identification of subsets of SIs that systematically impact FC sub-networks. 
We have further developed computationally efficient algorithms to extract the multi-level sub-networks from the full graphs and have proposed a tailored network-based inference frame to individually test each sub-network with multiple corrections based on permutation tests. Our method is also compatible with the existing methods aforementioned (e.g., PCA, CCA). For example, applying CCA to FCs and SIs in an extracted multi-level sub-network provides an estimate of association in the context of multiple regressions.   } 


{The contributions of this article are three-fold. First, we introduce MOAT, a novel method that can handle matrix-variate outcomes and vector-variate predictors. Compared to the existing models for multivariate outcomes and multivariate predictors \citep{zhuang2017family, wu2021multivariate, mihalik2022canonical, lu2023network}, MOAT can further account for the network structure within the matrix outcomes and between the outcome-predictor association patterns.  MOAT naturally prohibits most false positive associations because these associations are more likely distributed sparsely rather than gathered in organized sub-networks.
Secondly, we develop new algorithms to extract those multi-level sub-networks.
The computational load is low because we developed a tailored greedy peeling algorithm with multilinear complexity, making our approach compatible with the commonly used permutation tests that are often computationally intensive. 
Lastly, we proposed a novel network-level inference framework, where we utilize novel test statistics derived based on the multi-level dense subgraph properties in terms of size and density. This inference framework leads to a simultaneous enhancement of both sensitivity and specificity by leveraging graph combinatorial theories. 
}


The rest of this paper is organized as follows. In Section 2, we formally define the multi-level network structure and present how MOAT works in network extraction with the network-based inference method. In Section 3, we perform extensive simulation analyses for method validation
and comparison. In Section 4, we apply MOAT to a real structure-function neuroimaging dataset from the UK Biobank with 4,242 participants to systematically investigate the FC-SI associations. We conclude with discussions in Section 5.

\section{Our method}
\subsection{Data structure and problem set up}

We collect structural-functional neuroimaging data from independent subjects, indexed as $\{1,\ldots,D\}$. 
For each subject $d: 1\leq d\leq D$, we observe three sets of measurements: 
\begin{enumerate}[(i)]
    \item 
    \yuan{Independent variables: 
    a vector of $m$ SI measures $\boldsymbol{X}^{(d)}=
    \big(x_1^{(d)}, \ldots, x_m^{(d)}\big)^T$.
    This vector characterizes anatomical structures of the brain, such as white matter microstructure integrity measured by fractional anisotropy from DTI \citep{mori2008stereotaxic} and region-wise cortical thickness obtained from MRI \citep{tustison2014large}. 
    
    }

    \item Outcome variables: an adjacency matrix that stores pairwise FC measures $\boldsymbol{Y}^{(d)}_{n\times n}$ between $n$ brain regions. 
    Each element $y_{ij}^{(d)}$ of $\boldsymbol{Y}^{(d)}$ represents the strength of functional connection between brain regions $i$ and $j$ of subject $d$, calculated from functional imaging data such as resting state fMRI. 
    Thanks to the Brainetcome Atlas \citep{fan2016human}, researchers can align the FC brain region partitions across different participants, thus conveniently, their $\boldsymbol{Y}$ share a common node set.
    We model $\boldsymbol{Y}^{(d)}$ as the outcome variable due to the widely accepted view in neurology that brain structure determines neural functions \citep{buckner2008brain,bai2009abnormal,honey2010can}. 


    \item Confounding variables:  $\boldsymbol{\eta}^{(d)}=
    \big(\eta_1^{(d)}, \ldots, \eta_p^{(d)}\big)^T$. 
    These variables include profiling information such as age, sex, genetics, and environment that may potentially affect brain functional connectome in complicated ways. 
    
\end{enumerate}

\subsection{Multi-level graph representation}

We explore the brain structural-functional relationship by considering the following regression model: for each subject $d\in [1, D]$, 
\begin{align}
\label{gen_model}
    g(y_{ij}^{(d)})
    =&~
    \theta_{ij}^0
    +
    \sum_{k=1}^m
    \beta_{(ij),k}
    x_k^{(d)}
    +
    \sum_{p=1}^P
    \alpha_{ij}^p
    \eta_p^{(d)},
\end{align}
where $g(\cdot)$ is a link function,  $\theta_{i j}^0$ is the intercept,  $\beta_{(ij),k}$ is the coefficient of the SI measure $x_k$, and $\alpha_{i j}^p$ is the coefficient of the nuisance covariate $\eta_p$ \citep{zhang2023generalized}. 
The focal parameter of interest in the above regression model (\ref{gen_model}) is $\beta_{(ij),k}$, where a nonzero coefficient $\beta_{(ij),k}\neq 0$ signifies an association between an SI measure $x_k$ and the functional connection $y_{ij}$ between brain regions $i$ and $j$. Consequently, learning the set $\{\beta_{(ij),k}\neq 0 \}$ allows for the unveiling of brain-wide association patterns between SIs and FCs. 

\begin{figure}[!h]
 \vspace{-8pt}
\makebox[\textwidth][c]{
    \includegraphics[width=1\textwidth]{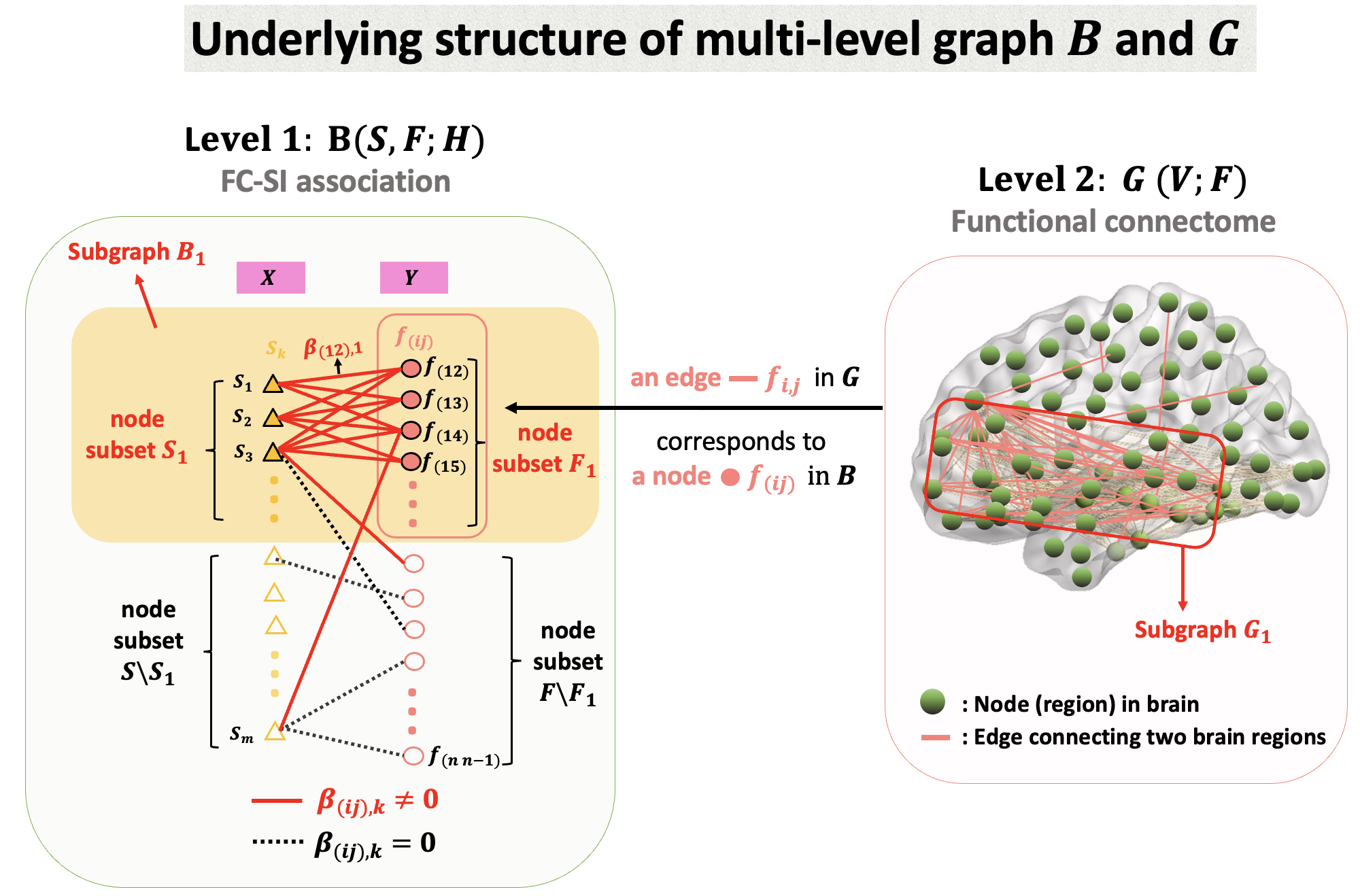}
}

\caption[A toy example of a multi-level graph $B$ and $G$, where there exists one FC-SI associated sub-network $B_1$]{\footnotesize \textit{
A toy example of a multi-level graph denoted as $B$ and $G$, where there exists one sub-network $B_1\in B$ exhibiting a significant association between FC and SI.  In Level 1,  $B=(S, F; H)$ is an input bipartite graph consisting of two node sets $S$ and $F$, representing the SI and FC measures respectively. The edge set $H$ is binary and determined by the regression coefficient $\boldsymbol{\beta}=\{\beta_{(ij),k}\}$ which infers the association between $y_{ij}$ (FC strength between brain regions $i$ and $j$) and $x_k$ ($k-$th SI measure).
In this toy example, there exists a sub-network $B_1=(S_1,F_1;H_1)\subset  B$ which contains SI-related edges $\{h_{(ij),k} | i,j,k \in B_1 \}$ with a much higher density than the rest $\{h_{(ij),k}| i,j,k \notin B_1 \}$. Given that the node set $F$ represents the whole-brain functional connectome, each node  $f_{(ij)} \in F$ comes from an edge in a brain connectome graph $G=\{V; F\}$ (Level 2 graph), where $V$ is the node set of brain regions, and $F$ is the edge set of interregional connections. $\{f_{(ij)}\} \in B_1$ are not isolated in the brain connectome; instead, they constitute an organized sub-graph $G_1\in G$.}
}
\label{tab:MOAT_graph_demo}
\end{figure}

\textbf{A multi-level graph model targeting $\{\beta_{(ij),k}\neq 0\}$ associations.} To facilitate downstream analysis, we let a matrix $\boldsymbol{\beta} = \{\beta_{(ij),k}\}_{\forall ijk} \in{\mathbb{R}^{\binom{n}{2} \times m}}$  to denote all SI-FC pair-wise associations. \citep{zhang2018tensor}.
We build the multi-level graph model based on the $\binom{n}{2} \times m$ matrix $ \boldsymbol{\beta} $. Specifically, at the first level, we define a bipartite graph $B=(S,F;H)$ to represent the matrix $\boldsymbol{\beta}$, where $S=\{1,\ldots,m\}$ (i.e., $|S|=m$) constitutes the node set of SI measures;
$F=\{1,\ldots,\binom{n}{2}\}$ (i.e., $|F|=\binom{n}{2}$) constitutes the node set of FC measures; 
and $H$ denotes the edge set.  Each element $h_{(ij),k}\in H$ signifies a non-zero association between FC and SI (i.e., $\beta_{(ij),k}\neq 0$).
We demonstrate the first level bipartite graph $B$ in the left panel of \autoref{tab:MOAT_graph_demo} . The second level of the multi-level graph model is a classic graph model reflecting the whole-brain connectome network, denoted as $G=(V;F)$, where $V$ is the node set of brain regions with size $|V|=n$, while $F$ is the edge set connecting brain regions with size $|F|\leq(n-1)/2$. Noticeably, each node $(i,j)$ in $F$ can also be interpreted as an edge in the brain functional connetome network $G(V;F)$. Thus, $F$ denotes both (i) the node set of the bipartite graph $B=(S,F;H)$ with $F=\{f_{(ij)} \}$, where $f_{(ij)}$ represents a node for the outcome $Y_{ij}$; (ii) the edge set of $G(V;F)$ with $F=\{f_{i,j} \}$, where $f_{i,j}=1$ indicates that brain areas $i$ and $j$ are connected.

In light of the highly organized brain structures and functions, it is neurobiologically sensible to model $\{\beta_{(ij),k}\neq 0\}$ in organized association patterns \citep{craddock2013imaging,bahrami2019analysis}. Specifically, we consider that a subset of brain structural predictors jointly influences connectome outcome variables within a functional subnetwork, which characterizes a plausible brain structure-function interaction \citep{zalesky2010network,cao2014topological}. Built upon this latent relationship pattern, we specify  that $\{\beta_{(ij),k}\neq 0\}$  predominantly concentrates within specific subgraphs denoted as $\{B_c \}$ where $c=1, \cdots, C$ and $B_c \subset B$. For simplicity,  we are going to illustrate the case where $c=1$ below.

We specify $B_1(S_1,F_1)$ as a doubly-dense multi-level subgraph (see dense graph studied in Tong MOAT, \cite{craddock2013imaging,wu2021multivariate}). At the first level, a subset of SI predictors of $\{X_k, k \in S_1\}$ condensely affect $\{Y_{ij}, ij \in F_1\}$:
\begin{equation}
\label{outer}
    \operatorname{Pr}\left(\beta_{(i j), k}\neq 0 \mid (ij)\in F_1, k \in S_1\right) 
    \gg 
    \operatorname{Pr}\left(\beta_{(i j), k} \neq 0 \mid (ij)\notin F_1 \textrm{ or } k \notin S_1\right). 
\end{equation}
At the second level, a connectomic subnetwork $G_1=(V_1; F_1)$ is an edge-induced sub-clique, where the edge subset of interest is $\{(i,j): (ij)\in F_1\}$. 
$G_1$ is also dense reflecting that SIs of $S_1$ are associated with connectomic edges in a network rather than sparsely and randomly distributed in the whole-brain connectome (ref Tong 2023). In \autoref{tab:MOAT_graph_demo}, we demonstrate a doubly-dense multi-level subgraph $B_1$ with red-bold edges. Provided with $\{B_c\}$, we can express the overall multi-level graph as (3) and (4) as follows:
\begin{align}
    & \text{Level 1 (bipartite) network: } B=\cup_{c=1}^C B_c \cup B_0, ~B_c=(S_c, F_c; H_c); \\
    & \text{Level 2 (unipartite) network: } G=\cup_{c=1}^C G_c \cup G_0, ~G_c=(V_c; F_c),
\end{align}
where $B_c$ and $G_c$ are dense subgraphs in $B$ and $G$ respectively, and $B_0$ and $G_0$ are the remaining graphs. 
Each node set $F_c$ in $B_c$ corresponds to a subset of edges in the functional connectome $G$, which induces one or  {multiple} cliques in $G$. For simplicity, we use $G_c$ to denote the clique(s) for the corresponding $B_c$.    
If $B$ is a random graph, then $B = B_0$ and for all $c=1,...,C$, $B_c=\emptyset$. Similarly, if $G$ is a random graph, then $G_c=\emptyset$. Otherwise, $G_c$ represents a connectome sub-network. 
In Figure \ref{tab:MOAT_graph_demo}, we demonstrate a graphical example of the multi-level structure of $B_c$ and $G_c$ when $c=1$. In summary, our multi-level network model assigns a small proportion of $\{\beta_{(ij),k}\neq 0\}$ to structured subnetworks reflecting systematic FC-SI association patterns. The patterns may not be captured by neither shrinkage regression models nor clustering/biclustering methods.

\subsection{Multi-level  subnetwork   estimation}
\label{MOAT_extraction}
In practice, neither $\{\beta_{(ij),k}\neq 0\}$ nor $\{B_c \}$ is known and it is challenging to simultaneously handle billions of FC-SI associations and estimate $\{B_c \}$ ($\{\beta_{(ij),k}\neq 0\}$) in one big model such as \eqref{gen_model} \citep{woo2014cluster, mbatchou2021computationally,marek2022reproducible}. 
To alleviate the computation burden, we take a divide-and-conquer approach and run one regression for each $k$, recognizing that both $\theta$ and $\alpha$ may also be different for each $k$. This strategy is commonly used in large-scale imaging and genetics data analysis \citep{zalesky2010network,schaid2018genome,chen2023identifying}.

Next, we extract the desired dense subgraphs $\{B_c\}$ based on $\boldsymbol{X}^{(d)}, \boldsymbol{Y}^{(d)}, \boldsymbol{\eta}^{(d)}$. 
Since $\{\beta_{(ij),k}\neq 0\}$ are unknown, we compute an inference measure $a_{(ij),k}$ as a surrogate to $\beta_{(ij),k}$:
each $a_{(ij),k}$ is produced by the statistical inference of a regression model for $x_k$ and $y_{ij}$. 
For example, $a_{(ij),k}$ can be the $-\log(p)$ for $\beta_{(ij),k}$, where $-\log(p)$ is a widely used metric in high-dimensional data analysis, such as 
Genome-wide association studies (GWAS) and neuroimaging analysis \citep{lasky2008genome,tang2016gapit,sun2022comparison}.
Now we propose the following criterion for selecting $(S_1,F_1)$:

\begin{equation}
\label{MOAT_obj}
    \underset{S_c \subseteq S, F_c \subseteq F, V_c \subseteq V}{\arg \max } \sum_{c=1}^C \frac{\sum_{k \in S_c, (i j)\in F_c} a_{(ij),k} }{\left(\left|S_c\right| \binom{|V_c|}{2}\right)^{\lambda_1 / 2}}+\frac{\sum_{i,j\in V_c, i<j} f_{i j}}{\left|V_c\right|^{\lambda_2}},
\end{equation}
where $\lambda_1, \lambda_2$ tune the impacts of the densities of $B_c$ and $G_c$, respectively. 
For example, when $\lambda_1=2$, the first term becomes the familiar quantity of subgraph density in network analysis.
We typically search $\lambda_1$ within the range of $(1,2)$.
Empirically, setting $\lambda_1=2$ usually forces $B_c$'s into singletons; while setting $\lambda_1$ below 1 often leads to sparse $B_c$'s. 
Likewise, we explore the parameter $\lambda_2$ within the same interval $(1,2)$. Deviating from this range for $\lambda_2$, either higher or lower, will yield results similar to those observed for $\lambda_1$.
Here, we follow the convention in neuroimaging analysis and select 
$\lambda_1$ and $\lambda_2$ using Kullback–Leibler (KL) divergence  \citep{johnson2001symmetrizing, yohai2008optimal,zhao2023mediation}. Detailed selection procedures are provided in Appendix A..

Directly solving \eqref{MOAT_obj} requires combinatorial computation.
Therefore, we propose a greedy peeling algorithm as a fast approximation. 
Our algorithm extends the greedy algorithm for single-level bipartite subgraphs extraction in \citet{wu2021multivariate} and \citet{chekuri2022densest}. 
We present a condensed version as Algorithm \ref{alg:MOAT_alg1} below, and relegate the detailed step-by-step algorithm to Appendix B.
For each multi-level subgraph  $B_c$ and $G_c$, Algorithm \ref{alg:MOAT_alg1} first initializes node sets $S_c$ and $F_c$ with the nodes $S$ and $F$ from the original full graphs, respectively. 
It then iteratively removes nodes with the smallest degree (say, $\tau\in S_c$ and $\phi\in F_c$) from either $S$ or $F$ (see Line 5 of the algorithm). 
At the end of each iteration $q$, the updated node set $F_c^{(q)}$ is used to construct the ``level 2" subgraph $G^{(q)}_c(V_c; F_c)$, and the corresponding output value of objective function \eqref{MOAT_obj} is recorded. This process of node removal and the construction of ``level 2" graph continues iteratively until all nodes have been excluded from $S_c$ or $F_c$, with the termination determined by whichever node subset is exhausted first. Ultimately, the algorithm returns the dense subgraph $B_c$ that maximizes \eqref{MOAT_obj} among all $\{B_c^{(q)}\}$ (see Line 14 of the algorithm).

The computational complexity of  Algorithm~\ref{alg:MOAT_alg1} is $\mathcal{O}\Big(\mathcal{C}|V|(|S|+|F|)\Big)$, where $\mathcal{C}$ depends on the number of the grid search, $|V|=n$,  $|S|=m$ and $|F|=\binom{n}{2}$ are the numbers of regions, SI measures, and FC measures, respectively.
Additionally, Theorem \ref{MOAT_th1} confirms the consistency of multi-level subgraph detection. 
In essence, the solution to the objective function \eqref{MOAT_obj} gives a consistent estimation of the true multi-level sub-network structure represented by ${B_c}$ (the set of edge-induced sub-networks). 
As the sample size $D\to\infty$, the likelihood of an incorrect edge assignment for ${B_c}$ approaches zero.

\begin{algorithm}[h!]
 \small
    \caption{Optimization of objective function (\ref{MOAT_obj})
    \label{alg:MOAT_alg1} - Condensed pseudo code}
    
    \begin{algorithmic}[1]
    \Statex \textbf{Input}: $B=(S,F;H)$, $G=(V,F)$, $\lambda_1$, $\lambda_2$ 
    \Statex \textbf{output}: $\{B_c\} $
    
        \State Define \textbf{function} [$B_c~$ density$(B_c)$]= greedy\_peeling\_MOAT [$B~ G~ \lambda_1~ \lambda_2]$:
       
            \State Initialize "level 1" subgraph nodes $S_c^1\leftarrow S, F_c^1 \leftarrow F$
            \For {$q=1,2,\ldots, m+\binom{n}{2}-1$}
                \State Let $\tau\in S_c^{(q)}$ and $\phi\in F_c^{(q)}$ be the nodes with smallest degree 

               \State \textbf{If}{$\sqrt{d} \tau \leq \frac{1}{\sqrt{d}} \phi $}, \textbf{then}
                remove $\tau$ from $S_c^{(q)}$;
                \textbf{otherwise},
                remove $\phi$ from $F_c^{(q)}$ 
            
                \State  Next, construct ``level 2" subgraph $G_c(V_c; F_c)$, based on current $F_c$

                \State Initialize "level 2" subgraph nodes $V_c^1 \leftarrow V$
                \For {$p=1,2,\ldots,n-1$}
                \State remove node with smallest degree from $V_c^{(p)}$, store each $V_c^{(p)}$ and the corresponding $F_c^{(p)}$
                 \EndFor
                 
                 \State Output $G^{(q)}_c(V_c; F_c)$ that maximizes 
                 $\frac{\sum_{i,j\in V_c, i<j} f_{i j}}{\left|V_c\right|^{\lambda_2}}$ among $\{V_c^{(p)}, F_c^{(p)}\}_{p=1}^{n-1}$
                 \State Replace $F_c^{(q)}$ with $F_c^{(p)}$ 
           
            \EndFor
            \State Output $B_c$ that maximizes (\ref{MOAT_obj})
            among $\{B_c^{(q)}\}_{q=1}^{m+\binom{n}{2}-1}$  
       
        \While {density$(B_c) >$ density($B$)}
            \State Fill median of $\{a_{(ij),k} | k \in S, (i j)\in F\}$  into $\{a_{(ij),k}|k \in S_c, (i j)\in F_c\}$, obtain updated $B^*$ and $G^*$
            
            \State [$B_c~$ density$(B_c)$]= greedy\_peeling\_MOAT [$B^*~ G^*~ \lambda_1~ \lambda_2~ ]$
        \EndWhile 

        \State Output all $\{B_c\} $
        \newline
        
    \end{algorithmic}
\end{algorithm}


\begin{customthm}{1}\label{MOAT_th1}
(\textbf{Consistency of subgraph detection}). let $\mathbf{U}^* \in \mathbb{R}^{|S|\times |F|}$ be a matrix storing the true edge membership in $B$, where each element $u_{(ij),k}^*=1$ if $\beta_{(ij),k}\neq0$, and $u_{(ij),k}^*=0$ otherwise. Similarly, let $\hat{\mathbf{U}} \in \mathbb{R}^{|S|\times |F|}$ store the edge membership estimated by optimizing ~\eqref{MOAT_obj}, where each element $\hat{u}_{(ij),k}=1$ if $x_k\in \hat{S}_c$ and $y_{ij}\in \hat{F}_c$; $\hat{u}_{(ij),k}=0$ otherwise. Then, for an arbitrarily small $\epsilon$, when the sample size $D\to \infty$, we have
\begin{align*}
    \mathbb{P}(||\mathbf{U}^*-\hat{\mathbf{U}}||_F<\epsilon) \to1,
\end{align*}
where $||.||_F$ denotes the frobenius norm.
\end{customthm}
The proof of Theorem \ref{MOAT_th1} is provided in Appendix~B.2.

\subsection{Reduced false positive findings by $B_c$ } \label{MOAT_section_graph}

Compared to methods that \textit{individually} select $\{\beta_{(ij),k}\}$ such as multiple testing approaches, our method selects nonzero $\beta$'s via \textit{dense} FC-SI associated sub-network $B_c$, which can drastically reduce false positive findings. 
Let $\{ \hat{\beta}_{(ij),k}\}$ denote the set of estimated association parameters from a sample; then $ \{ \hat{\beta}_{(ij),k} \neq 0 | \beta_{(ij),k} = 0\}$ indicate false positive findings. 
Following the common practice in neuroimaging and neurobiology \citep{margulis2000life}, we assume that false positive associations are randomly distributed in the brain space. 
The conventional approach using individual inference on $\{ \hat{\beta}_{(ij),k}\}$ may likely select many false positives $   \hat{\beta}_{(ij),k} \neq 0 | \beta_{(ij),k} = 0 $.
In contrast, our method returns few false positives.
The reason behind is demonstrated in the following lemma, which says false positives very rarely form dense subgraphs of moderate sizes.

\begin{lemma}
\label{MOAT_lemma1}
Assume that $B_c$ is observed from a random multi-level binary graph with a bipartite graph $B(S,F; H)$ in Level 1 and a unipartite graph $G(V; F)$ in Level 2. 
Suppose that $B_c$ is a multi-level subgraph that has:
(1) Edge density in $B_c$ with $\frac{\sum_{k \in S_c, (i j)\in F_c} I(\hat{\beta}_{(ij),k} \neq 0 | \beta_{(ij),k} = 0)}{\left|S_c\right|\left|F_c\right|} \geq {\gamma_1} \in\left(p_1, 1\right)$, where $p_1=\frac{\sum_{k \in S, (i j)\in F } I(\hat{\beta}_{(ij),k} \neq 0 | \beta_{(ij),k} = 0)}{|S||F|}$ is the proportion of false positive associations in $B$;
(2) Edge density in $G_c$  with $\frac{\sum_{i,j\in V_c, i<j} I(\hat{\beta}_{(ij),k} \neq 0 | \beta_{(ij),k} = 0)}{\binom{\left|V_c\right|}{2}} \geq {\gamma_2} \in\left(p_2, 1\right)$, where $p_2=\frac{\sum_{i,j\in V, i<j} I(\hat{\beta}_{(ij),k} \neq 0 | \beta_{(ij),k} = 0)}{\binom{|V|}{2}}$ is the proportion of false positive associations in $G$.
Furthermore, let $m_0, n_0=\Omega\left(\max \left\{m^\epsilon, n^\epsilon\right\}\right)$ for some $0<\epsilon<1$, where $\Omega$ denotes a loose lower bound. Then for sufficiently large $m,n$ with $\zeta(\gamma_1,p_1)m_0 \geq 4\log n(n-1)$,  $\zeta(\gamma_1,p_1)n_0(n_0-1) \geq 16\log m$, and $\zeta(\gamma_2,p_2)n_0 \geq 4\log n$, we have 
\begin{align}
\label{MOAT_lemma1_eq}
    & \mathbb{P}\left(\left|S_c\right| \geq m_0,\left|F_c\right| \geq \binom{n_0}{2}, \left|V_c\right| \geq  n_0 \right) \nonumber \\ 
    & \leq 2 m n^2 (n-1) \cdot \exp \left(-\frac{1}{8} \zeta\left(\gamma_1, p_1\right) m_0 n_0 (n_0-1) -\frac{1}{4} \zeta\left(\gamma_2, p_2\right) n_0^2\right),
\end{align}
where $\zeta(a, b)=\left\{\frac{1}{(a-b)^2}+\frac{1}{3(a-b)}\right\}^{-1}$. 
\end{lemma}


Lemma \ref{MOAT_lemma1} is proved in Appendix B.  
Lemma 1 states that the probability of identifying a multi-level subgraph of $\hat{B}_c$ composed of false positive associations $ \{ \hat{\beta}_{(ij),k} \neq 0 | \beta_{(ij),k} = 0\}$  
 exponentially converges to 0 as the sizes and densities of the multi-level sub-network increase.
In practice, the probability of a false positive network with reasonable size (e.g., $|S|\times |F|=10 \times 10$) and sound densities is less than $10^{-16}$.  
It is very unlikely that false positive FC-SI associations $ \{ \beta_{(ij),k} \neq 0 \}$ would form a large and dense subgraph $B_c$. 

\subsection{Inference for extracted \texorpdfstring{$\hat{B}_c$}{Lg}}
Recall from Section \ref{MOAT_extraction} that our study aims to identify specific subsets of SIs and FCs that exhibit systematic association patterns encoded by $B_c$. 
Performing Algorithm\autoref{alg:MOAT_alg1} returns a collection of such subgraphs $\hat{B}_c$. 
Our next goal is to conduct a network-level statistical inference to gauge the significance of each $\hat{B}_c$ with multiple corrections \citep{goeman2022cluster,zhang2023generalized,chen2023identifying}. 
Roughly speaking, we assess the statistical significance of each $B_c$ by testing:
\begin{align}
    \mathbb{H}_0&: \textrm{$B_c$ is not  a dense multi-level subgraph constituted by associated SI-FC pairs;} \nonumber \\
     \mathbb{H}_a&: \textrm{$B_c$ is a dense multi-level subgraph reflecting systematic FC-SI associations.} 
\end{align}
More precisely, under $\mathbb{H}_0$, the edges of $B_c$ is randomly distributed among all possible pairs. 
Per Lemma \ref{MOAT_lemma1}, it is rare to observe large and dense multi-level subgraph $B_c$ under the null. 
Therefore, we can straightforwardly perform the commonly used permutation testing strategy in neuroimaging statistics to assess the significance of $\hat{B}_c$ while controlling the FWER (\cite{zalesky2010network,nichols2012multiple,woo2014cluster}. 
However, our testing object is a multi-level subgraph, which is different from the voxel-based ``clusters'' commonly encountered in conventional cluster-extent inference because the rareness of  $\hat{B}_c$ is jointly determined by both the densities and sizes of the dense bipartite and clique of $\hat{B}_c$ instead of a measure of cluster-extent (e.g., the number of voxels).
To address this challenge, we propose a novel test statistic $\mathcal{T}(\hat{B}_c)$, proportional to the upper bound of the probability of observing a clique of certain size and density in a random graph, which appeared in \eqref{MOAT_lemma1_eq}: 



\begin{align}
    \mathcal{T}(\hat{B}_c)=\exp \left(-\frac{1}{4} \zeta\left(\gamma_1, p_1\right) |S_c||F_c| -\frac{1}{4} \zeta\left(\gamma_2, p_2\right) |V_c|^2\right),
\end{align}
where $\zeta(a, b)=\left\{\frac{1}{(a-b)^2}+\frac{1}{3(a-b)}\right\}^{-1}$, 
$\gamma_1=\frac{\left|H_c\right|}{\left|S_c\right|\left|F_c\right|}$,
$p_1=\frac{\sum_{k \in S, (i j)\in F} I(\hat{\beta}_{(ij),k} \neq 0)}{|S||F|}$,
$\gamma_2=\frac{\left|F_c\right|}{\binom{|V_c|}{2}}$,
$p_2=\frac{\sum_{i,j\in 
V, i<j} I(\hat{\beta}_{(ij),k} \neq 0)}{\binom{|V|}{2}}$.
We formally present our proposed \textit{network-based permutation test} for the significance of each extracted $\hat{B}_c$ in Algorithm \ref{alg:MOAT_alg2}.
The permutation procedure outlined in Algorithm \ref{alg:MOAT_alg2} is effective in simulating the null distribution of the test statistic $T(\hat{B}_c)$. Therefore, FWER can be controlled effectively, yielding a corrected $p$-value for each extracted $\hat{B}_c$.
\begin{algorithm}[H]
 \small
    \caption{Assess the significance of $\{\hat{B}_c\}_{c=1,\ldots,C}$}
    \label{alg:MOAT_alg2}
    
    \begin{algorithmic}
    \Statex \textbf{Input}: $\{a_{(ij),k}\},\hat{C}\geq1,\{\hat{B}_c\}, \alpha$  
    \Statex \textbf{output}: FWER-controlled significance values for each $\hat{B}_c$

    \State 1. Choose a sound cut-off $\hat{r}$ and binarize graph $B[\hat{r}]$:$B[\hat{r}]_{(ij),k}=I(a_{(ij),k}>\hat{r})$
    \State 2. Estimate edge densities for $B$ and $G$, respectively:
    \begin{align*}
        &\text{Overall density}: \hat{p}_1=\frac{\sum_{k\in S, (ij)\in F} I(a_{(ij),k}>\hat{r})}{|S||F|},~~ 
        \hat{p}_2=\frac{\sum_{i,j\in V, i<j} f_{ij}}{\binom{|V|}{2}};
        \\ 
        & \text{Within-subgraph density}:
        \hat{\gamma}_1=\frac{\sum_{k \in \hat{S}_c, (i j)\in \hat{F}_c } I(a_{(ij),k}>\hat{r})} {|\hat{S}_c||\hat{F}_c|},~~ 
        \hat{\gamma}_2=\frac{\sum_{i,j\in \hat{V}_c, i<j }f_{ij}} {\binom{|\hat{V}_c|}{2}}.    
    \end{align*}

    \State 3. Calculate test statistic $\mathcal{T}_0(\hat{B}_c)$ for the current $\hat{B}_c$ 

    \State 4. Shuffle group labels of data $P$ times and implement MOAT on each shuffled graph

    \State 5. Store the maximal test statistic for each simulation $l=1,\ldots, L:$
    \begin{align*}
    \mathcal{T}_l=\sup _{c=1, \ldots, \hat{C}^l}\exp \left(-\frac{1}{4} \zeta\left(\gamma_1^l, p_1^l\right) |\hat{S}^l_c||\hat{F}^l_c| -\frac{1}{4} \zeta \left(\gamma_2^l, p_2^l\right) |\hat{V}^l_c|^2 \right)
\end{align*}

    \State 6. Calculate the percentile of $\mathcal{T}_0(\hat{B}_c)$ in $\{\mathcal{T}_l\}_{l\in(1,\ldots,L)}$ as the FWER $q$-value and reject $\mathbb{H}_0$ if $q<\alpha$, where $\alpha$ is a pre-specified significant level
        
    \end{algorithmic}
\end{algorithm}


\textit{{Evaluating the joint effect of multiple SIs on FCs.}}
With each $\hat{B}_c=(S_c, F_c)$, we have a set of structural measures $\{ X_k, k \in S_c\}$ 
 associated with functional measures $\{ Y_{ij}, (ij)) \in F_c\}$. However, this does not automatically provide the joint effect (i.e., $\sum_{(ij)) \in F_c, k \in S_c} \beta_{(ij),k} X_k$) of the selected SIs on each selected FC measure. To assess the joint effect, we can adopt the existing multivariate-to-multivarite analysis tools (e.g., CCA). 
Detailed procedures for applying CCA on $\hat{B}_c$ is provided in Appendix D.
  Alternatively, one can conduct low-rank regression on outcomes and predictors in each $ \hat{B}_c$ to estimate the final effect size \citep{vounou2010discovering, wang2012identifying, zhu2014bayesian, kong2019l2rm}.
 

\section{Simulation}
In this simulation study, we probed whether MOAT can extract informative subgraphs $\{\hat{B}_c\}$ from the multi-level graph $B$ and $G$ with high accuracy and replicability. 
We evaluate MOAT to finite-sample simulation data under various conditions (e.g., different sample sizes and effect sizes) with comparisons to several commonly used biclustering methods and sCCA-based methods.

\subsection{Synthetic data}
 We generate synthetic FC data $\mathbf{Y}^{(d)}=\Big\{y^{(d)}_{ij}\Big\}_{i,j<n}$ and SI data $\boldsymbol{X}^{(d)}=\big(x_1^{(d)}, \ldots, x_m^{(d)}\big)^T$ based on the following multivariate Gaussian distribution: 
   \begin{align}
   \label{MOAT_simudata}
    \begin{pmatrix}\boldsymbol{X}^{(d)}\\
    \mathbf{Y}^{(d)}
    \end{pmatrix} &\sim  N
    \begin{bmatrix}
    \begin{pmatrix}
    \boldsymbol{\mu}_X\\
    \boldsymbol{\mu}_Y
    \end{pmatrix}\!\!,&
    \left(\begin{array}{cc} \mathbf{\Sigma}_{X,X} & \mathbf{\Sigma}_{X,Y}\\ \mathbf{\Sigma}_{Y,X} & \mathbf{\Sigma}_{Y,Y} \end{array}\right)
    \end{bmatrix},
   \end{align} 
where $\begin{pmatrix}
    \boldsymbol{\mu}_X\\
    \boldsymbol{\mu}_Y
    \end{pmatrix}$ is the partitioned mean vector of SI and FC data respectively, and $\mathbf{\Sigma}=\left(\begin{array}{cc} \mathbf{\Sigma}_{X,X} & \mathbf{\Sigma}_{X,Y}\\ \mathbf{\Sigma}_{Y,X} & \mathbf{\Sigma}_{Y,Y} \end{array}\right)$ is the partitioned variance-covariance matrix.
For simplicity, we set $\begin{pmatrix}
    \boldsymbol{\mu}_X\\
    \boldsymbol{\mu}_Y
    \end{pmatrix}$ as a zero vector, representing normalized data, while the construction of $\mathbf{\Sigma}$ depends on two key factors: the multi-level network structure and effect sizes (i.e., FC-SI association strength). Both of these factors are elaborated upon in the following paragraph.

\begin{figure}[p]
\makebox[\textwidth][c]{
    \includegraphics[width=0.9\textwidth]{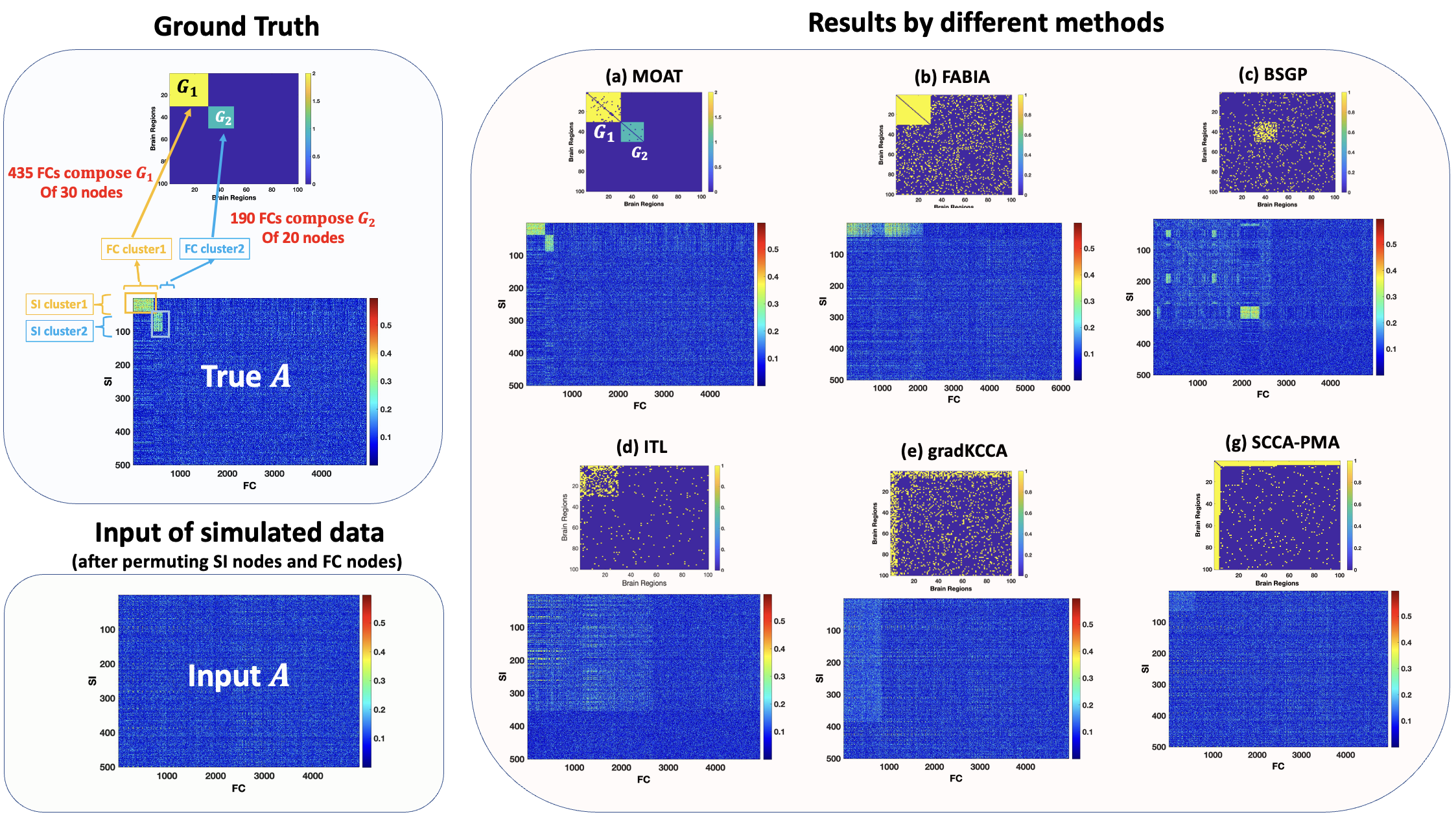}
}
\vspace{-15pt}
\caption[Applying MOAT and comparative methods to synthetic data]{\footnotesize \textit{Applying MOAT and comparative methods to an inference matrix $\boldsymbol{A}_{500 \times 4950}$ with sample size $D=200$. $\boldsymbol{A}_{500 \times 4950}$ is simulated to store partial correlations between $500$ SI measures and $4950$ FC measures, where the $4950$ FC measures are obtained from region-wise functional connections between 100 brain regions. Two sub-networks $B_1$ and $B_2$ with FC-SI partial correlations $(\rho_1,\rho_2)=(0.40,0.35)$ are generated within $B$, while the partial correlations for the rest of $B$ are set to be $\rho_0=0.15$. 
The right panel shows the results of applying MOAT, biclustering and sCCA methods.
Both MOAT and biclustering methods are more accurate in revealing sub-network patterns than sCCA methods since they incorporate cluster/network information. When jointly evaluating TPR and TNR, MOAT outperforms other methods, as TPR and TNR rely heavily on accurate sub-network extraction and inference.  }}
\label{tab:MOAT_simu_2d}
\end{figure}

In this simulation, we consider $500$ SIs and $4950$ FCs, where the FC measures are calculated based on a brain network with $100$ regions, resulting in ${100\choose 2}=4950$ pairwise connectivity values.  To determine the network structure for $\mathbf{\Sigma}$, we consider the following multi-level graph consisting of (i) a bipartite graph $B=(S,F;H)$ depicting the FC-SI associations, where $|S|=500, |F|=4950$; (ii) a unipartite graph $G=(V;E)$ depicting the brain functional connectome, where $|V|=100$. Specially, we generate two sub-networks within $B$, denoted as $B_1$ and $B_2$, characterized by higher FC-SI partial correlations $\rho_1$ and $\rho_2$ than the rest of $B$. $B_1$ consists of $40$ SI measures and $435$ FC measures, where the $435$ FC measures collectively compose a functional connectome $G_1$ of $30$ brain regions; $B_2$ consists of $60$ SI measures and $190$ FC measures, where the $190$ FC measures collectively compose another functional connectome $G_2$ of $20$ brain regions. For a visual representation of these two sub-networks, please refer to the graph illustration in \autoref{tab:MOAT_simu_2d}.

\setlength{\intextsep}{0pt}

Built on this network architecture, we configure the covariance matrix $\mathbf{\Sigma}$ such that $\rho_1,\rho_2>\rho_0$ to emulate different effect sizes. Here, we set $\rho_0=0.15$ as the partial correlation of FC-SI edges outside of $B_1$ and $B_2$. Next, by correlating the FC and SI data simulated from (\ref{MOAT_simudata}) using the aforespecified $\begin{pmatrix}
    \boldsymbol{\mu}_X\\
    \boldsymbol{\mu}_Y
    \end{pmatrix}$ and $\mathbf{\Sigma}$, we obtain an FC-SI association  matrix $\boldsymbol{A}_{500\times4950}$.
$\boldsymbol{A}$ governs the edge variable $H$ in the  bipartite graph $B=(S,F;H)$ by $h_{(ij),k}=I(a_{(ij),k}>r)$, where $r$ is a pre-selected threshold for correlation strength. Lastly, to assess MOAT performance under different settings, 
three configurations of $(\rho_0,\rho_1,\rho_2; D)$ are simulated: $(0.15, 0.55,0.60; 200)$, $(0.15, 0.60,0.45; 300)$, and $(0.15, 0.70,0.40; 400)$, where $D$ represents the sample size as defined previously. For each configuration, we simulate 500 repeated data sets $\{\boldsymbol{A}^l\}_{l\in(1,\ldots,500)}$ to better access accuracy and replicability of MOAT. 

\captionsetup{labelformat=empty}
\begin{figure}[!]
\makebox[\textwidth][c]{
    \includegraphics[width=0.9\textwidth]{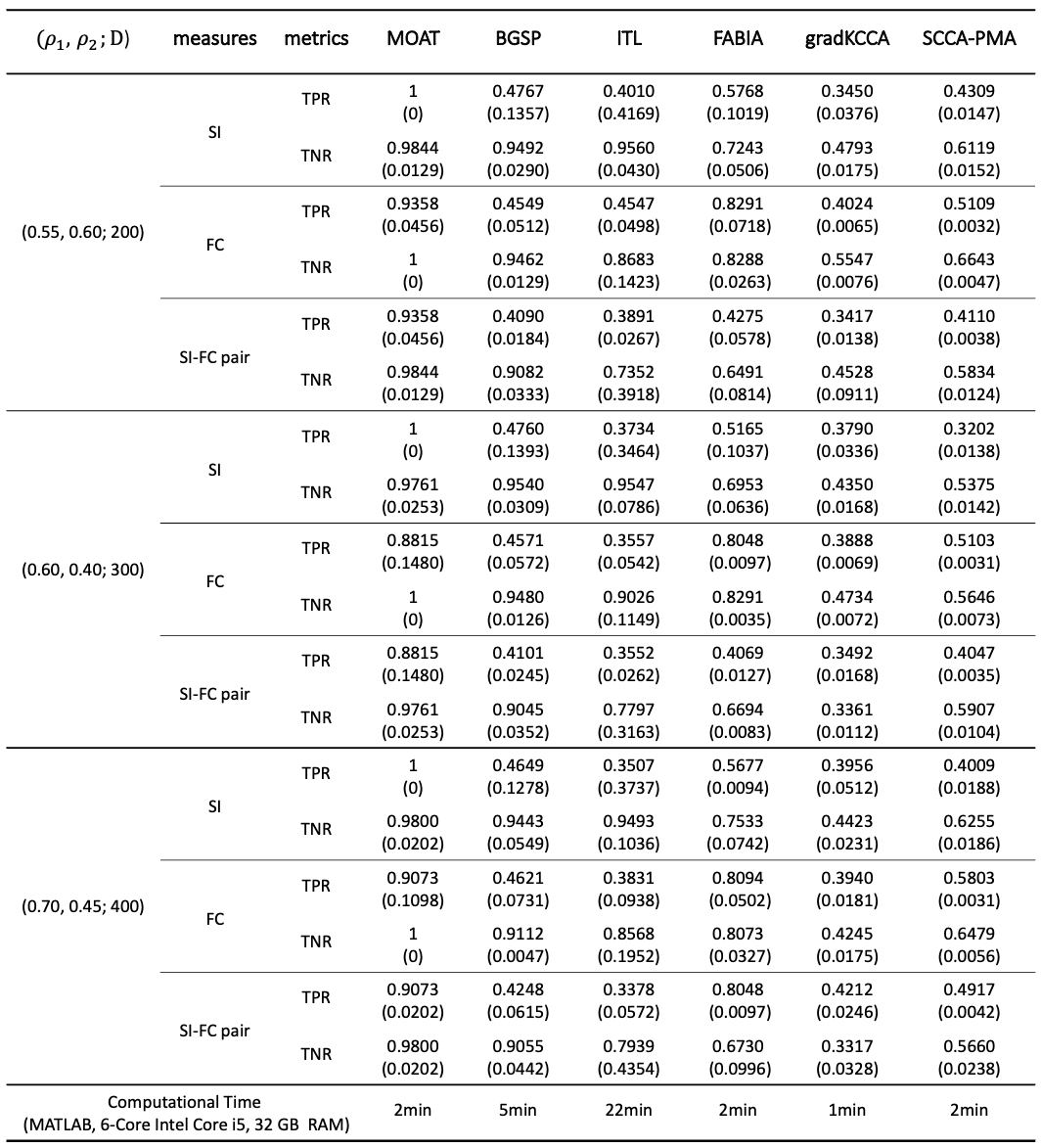}
}
\caption[Inference results of MOAT and comparative methods under different settings]{\footnotesize Table1: \textit{ Inference results of MOAT and comparative methods under different settings. $\rho_1,\rho_2$ denote the partial correlations between SI and FC measures in the dense sub-networks $B_1$ and $B_2$, respectively, and $D$ denotes the sample size. 
We summarize the means (standard deviations) of the TPR and TNR for identified SIs, FCs, and FC-SI pairs based on 500 repeated simulations.
The results show that MOAT outperforms the compared methods under different scenarios, indicating its effectiveness in identifying the correct dense sub-networks with true positive FC-SI associated pairs.
}}
\label{fig:MOAT_simu_table}
\end{figure}
\captionsetup[figure]{labelformat=default}

\subsection{Performance evaluation}
For each simulated dataset, we apply MOAT to estimate the multi-level sub-networks $\hat{B}_c$ containing strong FC-SI associations and perform our proposed network-based permutation test outlined in Algorithm \autoref{alg:MOAT_alg2} on $\boldsymbol{A}^l$. Regarding $ {B}_c$ extraction and $\{\beta_{(ij),k} \neq 0\}$ identification, we benchmark MOAT against a few popular appoaches including (i) three biclustering methods that are commonly used for sub-network detection: Bipartite Spectral Graph Partitioning (BSGP) \citep{wieling2009bipartite}, Information Theoretic Learning (ITL) \citep{erdogmus2002information}, and Factor Analysis for Bicluster Information Acquisition (FABIA) \citep{hochreiter2010fabia}; (ii) two sCCA-based methods that identify and measure the associations between two canonical/latent types of variables: a Large-Scale Sparse Kernel Canonical Correlation method proposed by \cite{uurtio2019large}, and sCCA through a penalized matrix decomposition (sCCA-PMA) proposed by \cite{witten2009penalized}. 

We evaluate methods' performance by assessing the deviation of the estimated $\hat{B}_c$ from true $B_c$ at both node-level, and edge-level (i.e., $\hat{\beta}_{(ij),k} \neq 0$ v.s true $\beta_{(ij),k} \neq 0$). Specifically, we consider the comparions  from the following three perspectives: SI variable selection, FC variable selection, and FC-SI pair selection. We use true positive rate (TPR) and true negative rate (TNR) as the evaluation criteria for both node-level and edge-level deviations. TPR is determined by the proportion of FC/SI nodes or FC-SI edges in $B_c$ that can be recovered by $\hat{B}_c$; TNR is determined by the proportion of FC/SI nodes or FC-SI edges in $B/B_c$ that can be recovered by $B/\hat{B}_c$.



\autoref{tab:MOAT_simu_2d} provides a graphical overview of the performance of each method.
Table \hyperref[fig:simu_table]{1} demonstrates the performance of all methods under multiple settings. The TPR and TNR are determined by the accuracy of both sub-network extraction and network-level inference.  In general, both MOAT and biclustering-based methods can recover sub-network patterns more accurately than sCCA-based methods because the network structures of FC-SI association patterns can be better recognized. 
Under different settings, MOAT can detect the target sub-networks with high sensitivity with few or none false-positive FC-SI edges because the cost of removing a true positive association or including a false positive edge is very high, as regulated by objective function (\ref{MOAT_obj}). The performance of biclustering methods is also improved with increased effect sizes with low false positive rates and medium to low sensitivity. In contrast, sCCA-based methods is invariant to different effect sizes, and may miss the underlying FC-SI association patterns due to various noise.  

Overall, MOAT is robust to noise and sensitive to organized FC-SI association patterns. MOAT outperforms comparable biclustering and sCCA methods under different settings, especially when systematic FC-SI association patterns are present. This superiority stems from MOAT's ability to accurately extract FC-SI association patterns through multi-level sub-network analysis and tailored sub-network-level inference.


\section{Study of FC-SI associations in brain connectome data}
\subsection{UK Biobank sample and neuroimaging data}

We aim to investigate the systematic effects of certain structural brain imaging measures on the functional connectome using  UK Biobank data \citep{sudlow2015uk}. The UK Biobank is a vast biomedical database with approximately half a million participants from the UK, where a total of 40,923 healthy individuals were found to have usable resting-state fMRI (rs-fMRI) data that passed quality control \citep{alfaro2018image}. Among them, a subgroup of 4,242 individuals possessed complete data on 
the following three sets of measurements we have chosen to focus on in this study:

\begin{enumerate}[(i)]
    \item      
    {\textit{105 SI measures}: we collected 105 SI variables including 39 white matter integrity measures and 66 cortical thickness measures. 
    The white matter integrity reflects the overall health and coherence of brain white matter and was assessed by fractional anisotropy (FA) obtained from DTI data in this study. The DTI data was pre-processed using ENIGMA DTI protocols \citep{jahanshad2013multi} and white matter tracts were labeled based on the JHU ICBM DTI-81 Atlas \citep{smith2006tract, mori2008stereotaxic}. A complete list of the 39 regional white matter tracts can be found in Appendix C.3.
    On the other hand, cortical thickness measures gauge the width of the gray matter of the human cortex, and were obtained from T1 MRI and labeled based on the FreeSurfer atlas \citep{tustison2014large}.}
    

    \item 
     {\textit{30,135 FC measures}: functional connectome data were obtained from rs-fMRI data based on Brainnetome Atlas \citep{fan2016human}. We first performed rs-fMRI preprocessing for all participants and then extracted the averaged time series of blood-oxygen-level-dependent (BOLD) signals from 246 functional brain regions, resulting in $\binom{246}{2}=30,135$ region-pair FC measures. Details of imaging acquisition and fMRI preprocessing are provided in Appendix C.1. }
    

    \item 
    \textit{4 confounding variables}: we adjusted four  confounding variables including age (years: $61.46 \pm 7.40$), sex (M/F: 2003/2239), educational level (years: $17.37 \pm 3.92$), and body mass index (BMI) ($kg/m^2: 26.35 \pm 4.30$). These variables have been used in previous neuroimaging literature on studying brain functional connectivity \citep{miller2016multimodal, alfaro2021confound, bischof2015obesity,agusti2018interplay}.  
\end{enumerate}

 



\subsection{Results}

We applied MOAT to the multimodal imaging data from the qualified 4,242 UK Biobank participants. First, we obtained the FC-SI association inference matrix $\boldsymbol{A}_{105 \times 30135}$. Each entry in $\boldsymbol{A}$ is $a_{(ij),k}=-\log (p_{(ij),k})$, where $p_{(ij),k}$ represents the $p$-value testing the association between the $k$-th SI measure and the FC outcome between two brain regions $i$ and $j$. 
Next, we performed a hard-thresholding sparsity constraint by setting $a_{(ij),k}=a_{(ij),k} I(a_{(ij),k}<\epsilon)$ for
some positive integer $\epsilon$ \citep{zhang2023generalized}.
We then applied our proposed greedy peeling algorithm \ref{alg:MOAT_alg1} to the inference matrix $\boldsymbol{A}$, with tuning parameters $\lambda_1=1.25, \lambda_2=1.5$  selected by the KL divergence with a mixed Bernoulli distribution based on random graphs $B$ and $G$. Algorithm \ref{alg:MOAT_alg1} returned one multi-level sub-network $\hat{B_1}\in B$. 
Lastly, we performed the network-level statistical inference on $\hat{B_1}$ using Algorithm \ref{alg:MOAT_alg2}. The testing results showed that the systematic association pattern of $\hat{B_1}$ is statistically significant ($p<0.0001$).

\begin{figure}[!h]
\makebox[\textwidth][c]{
    \includegraphics[width=1\textwidth]{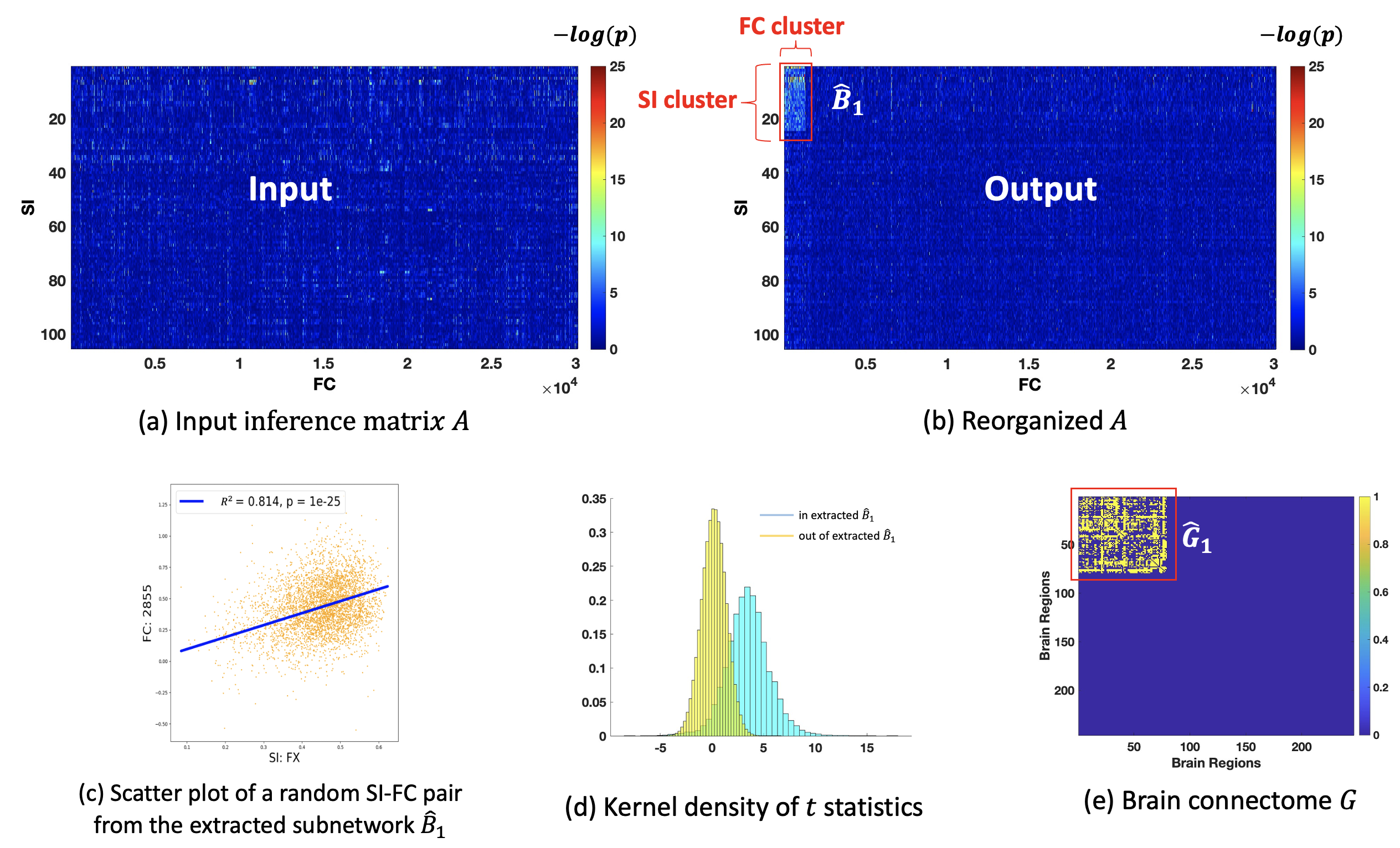}
}
\caption[Application of MOAT on a real neuroimaging dataset obtained from the UK Biobank.]{\footnotesize \textit{Application of MOAT on a real neuroimaging dataset obtained from the UK Biobank.  (a) shows a heatmap of the inference matrix $\boldsymbol{A}_{105 \times 30135}$. Each element is a $-log(p)$ value testing the association between SI and FC measures. (b) shows a heatmap of a reorganized $\boldsymbol{A}$ with element in the detected subgraph $\hat{B}_1$ pushed to the top left corner. Within the second level $G$ composed by the selected FCs, there exists an organized subgraph $G_1\subset G$ revealed in (e). As shown in (c) and (d), the FC-SI pairs within the identified $\hat{B}_1$ have significantly stronger associations compared to those outside of the network. }}
\label{tab:MOAT_real_2d}
\end{figure}

Specifically, results show that $\hat{B_1}$ comprised $|S_1|=23$ SI measures and $|F_1|=1316$ FC outcomes, as highlighted in \autoref{tab:MOAT_real_2d}(b). Furthermore, the extracted $|F_1|$ unfolded into a dense clique $\hat{G}_1 \in G$ consisting of $|V_1|=79$ regions, as illustrated in \autoref{tab:MOAT_real_2d}(e). 
The FC-SI pairs within the identified sub-network $\hat{B}_1$ demonstrate significantly stronger associations compared to those outside of the network,  as evidenced by the high $R^2$ and $t$-statistics shown in \autoref{tab:MOAT_real_2d} (c-d).
The 23 extracted SI measures consist of 3 cortical thickness measures and 20 FAs: the three cortical thickness measures correspond to the mean thickness of the parahippocampal, superior temporal, and cuneus gyrus; while for the 20 FA measures extracted, the top four with the strongest FC associations are CST-R (corticospinal tract, right hemisphere), CST-L (corticospinal tract, left hemisphere), ICP (inferior cerebellar peduncles), and FX (fornix). More detailed information about the remaining 16 FA measures can be found in Appendix E.3. \autoref{tab:MOAT_real3d} (left panel) illustrates the names and spatial locations of the 20 selected FAs.

The right panel in \autoref{tab:MOAT_real3d} shows the spatial distributions of within-$\hat{B}_1$ brain regions (79 regions in total), where they are predominantly located in six cortices: frontal, subcortical, temporal, parietal, insular, and limbic. Moreover, these regions consist of several well-defined brain functional networks including  temporo-frontal, somatomotor, ventral attention, frontoparietal, and (partial) default mode network (DMN). Overall, \autoref{tab:MOAT_real3d} provides a 3D demonstration showcasing the systematic association patterns between the subsets of SIs and FCs revealed by MOAT. Notably, both the FC-SI significantly associated sub-network ($\hat{B}_1$) and the brain functional sub-connectome ($\hat{G}_1$) exhibit well-organized topological structures. 




\begin{figure}[tb!]
\makebox[\textwidth][c]{
    \includegraphics[width=1\textwidth]{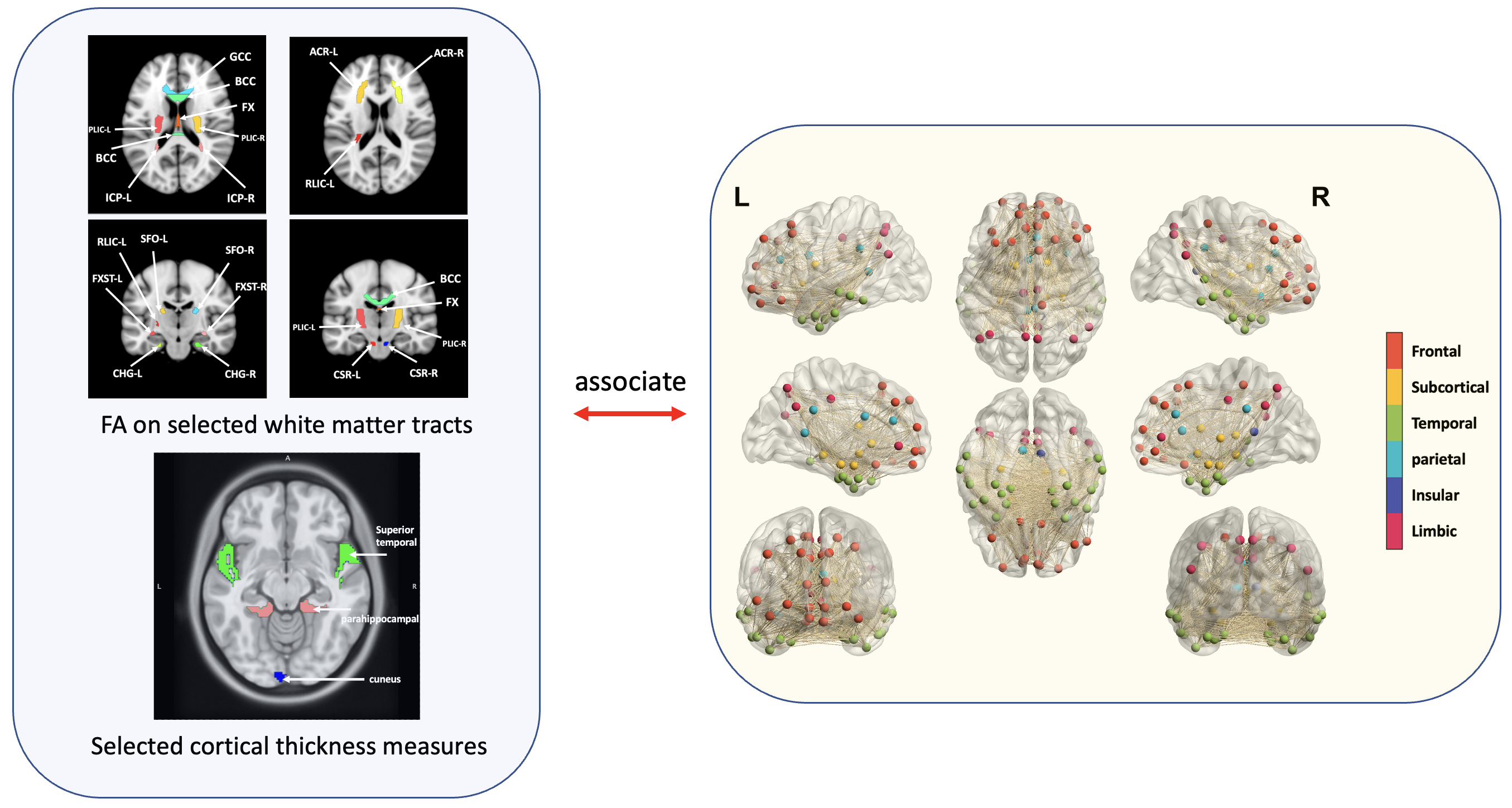}
}
\caption[Extracted FC-SI associated sub-networks by MOAT]{\footnotesize \textit{Extracted FC-SI associated sub-networks by MOAT, adjusted for age, sex, educational level, and BMI. Specifically, the left panel shows the SI cluster in the extracted sub-networks, which contains 20 FA measures and 3 cortical thickness measures. The top four FA measures with remarkably stronger  associations with the detected FC sub-network are CST-R, CST-L, ICP, and FX. The three selected cortical thickness measures correspond to the mean thickness of the parahippocampal, superior temporal, and cuneus gyrus. The right panel shows the detected FC sub-network that is strongly impacted by the 23 selected SIs. The detected FC sub-network covers 79 brain regions and composes several high-level cognitive brain networks including the default mode network, temporo-frontal network, somatomotor, ventral attention, and frontoparietal network. }}
\label{tab:MOAT_real3d}
\end{figure}

\captionsetup[figure]{labelformat=empty}
\captionsetup{labelformat=empty}
\begin{figure}[!]
\makebox[\textwidth][c]{
    \includegraphics[width=1\textwidth]{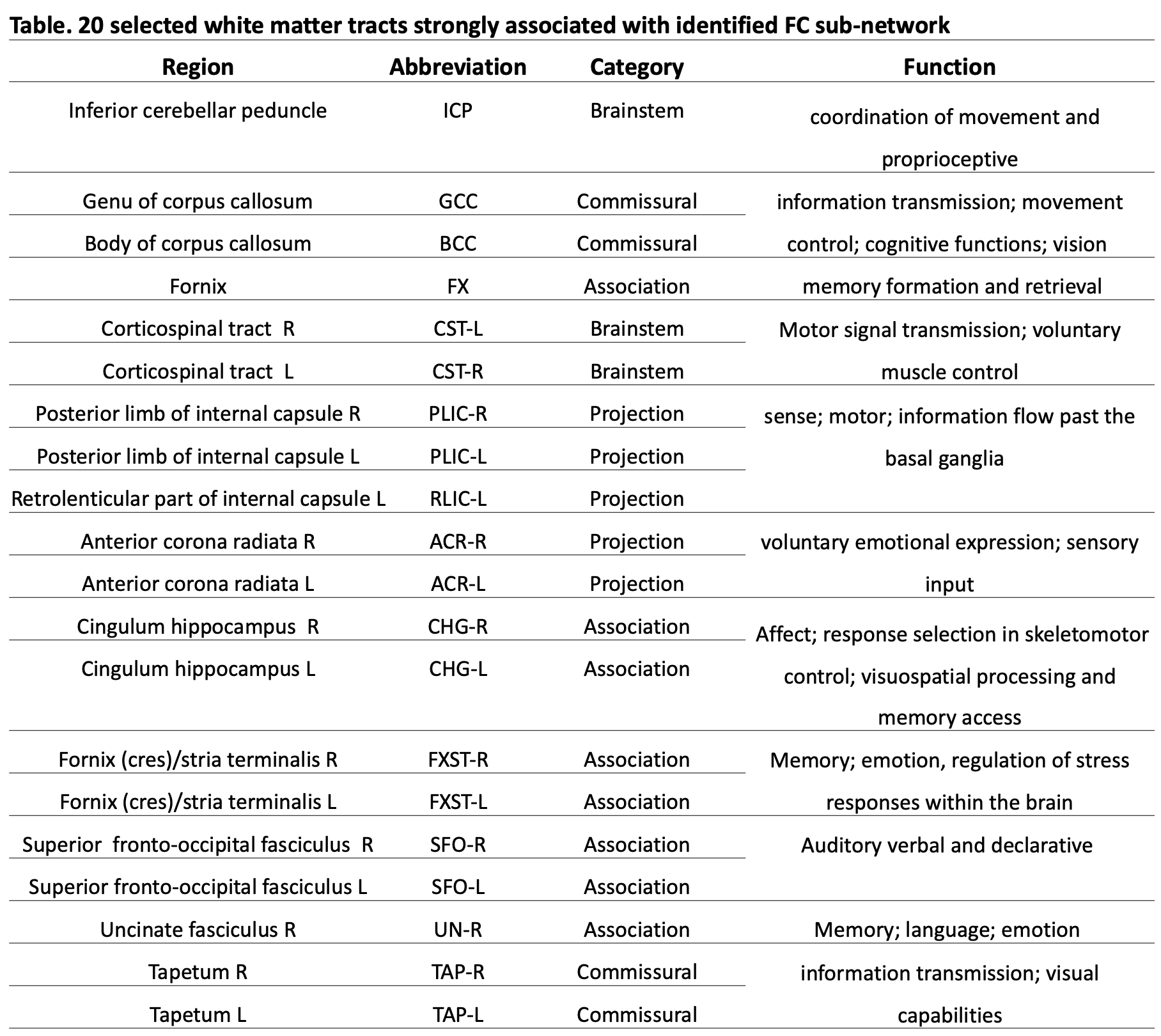}
}

\caption[20 selected white matter tracts strongly associated with identified FC sub-network]{\footnotesize Table 2: \textit{ 20 selected white matter tracts strongly associated with identified FC sub-network
}}
\label{fig:MOAT_FAs_20selected}
\end{figure}

We further applied CCA on the extracted sub-network $\hat{B}_1$ to quantitatively measure the canonical associations among the FC-SI pairs within $\hat{B}_1$. Results showed that the sample canonical correlations of the first three canonical variate pair in  $\hat{B}_1$ were $0.81$, $0.69$, and $0.68$ respectively. 
In contrast, we performed sparse CCA proposed by \cite{witten2009penalized} on the full graph $B$ and $G$, given the ultra high dimensionality of data. This yielded sample canonical correlations of $0.18$, $0.15$, and $0.14$ for the first three canonical variate pairs, respectively.
Notably, MOAT can better recognize the underlying large-scale FC-SI association patterns and then provide an improved estimation of the multivariate-to-multivariate association.


In summary, the application of MOAT helps to unfold the complex yet systematical and strong interplay between subsets of structural and functional measures of the human brain. Our findings suggest i) FC-SI associations are highly concentrated in a subset of SIs and FC sub-networks rather than exhibiting a whole-brain diffuse distribution pattern; ii) several FC sub-networks are primarily influenced by white matter integrity measures (refer to Table 2 in Appendix E.3 for possible mapping relationships); iii) multiple SI measures jointly affect the overall FC outcomes based on MOAT-guided CCA analysis.  While, on a high level, our results align well with previous medical findings \citep{cheung2008diffusion,pradat2009biomarkers, chaddock2013white,corrigan2015brainspotting}, MOAT reveals more refined patterns with improved spatial specificity and biological interpretability.   

\section{Discussion}
{Our newly developed approach, MOAT, offers a novel strategy to investigate the complex association patterns between multimodal neuroimaging data with matrix-outcomes (FCs) and a vector of imaging predictors (SIs). MOAT deciphers the complex FC-SI association patterns in a multi-level graph structure revealing the joint effect of a small set of SI predictors on FC sub-networks. The multi-level graph structure can effectively reduce the number of parameters while preserving the spatial specificity of FCs and SIs. MOAT delivers findings in organized multi-level sub-networks largely suppressing individual false positive FC-SI associations (see Lemma \ref{MOAT_lemma1} in section \ref{MOAT_section_graph}).  We developed computationally efficient algorithms to extract multi-level sub-networks. We further showed the consistency of the MOAT method. In addition, we develop a tailored network-level inference approach to test the extracted multi-level sub-networks while controlling FWER. Last, MOAT is also compatible with existing multivariate-to-multivariate analysis tools (e.g., CCA).  }

{In our case study, we investigated the FC-SI associations based on a large sample and revealed systematic association patterns with neurological explanations. This may enhance our understanding of how the brain structure and function interactively work during resting states and may lead to insights that can guide future cognitive and psychiatric therapy. However, since UK biobank participants mainly consist of elder Caucasians, our conclusion may be limited. Further investigation and integrated analysis is required to gain more comprehensive understanding of the FC-SI associations. The software package for MOAT is available at \href{https://github.com/TongLu-bit/MultilayerNetworks-MOAT}{https://github.com/TongLu-bit/MultilayerNetworks-MOAT}.}

\hfill \break
\textbf{Declaration of interest}: none.

\hfill \break
\textbf{Acknowledgments}

Tong Lu and Shuo Chen were supported by the National Institutes of Health under Award Numbers 1DP1DA04896801, EB008432, and EB008281.
Yuan Zhang was supported by the National Science Foundation under Award Number DMS-2311109.

\bibliographystyle{apalike}
\bibliography{Reference}

\end{sloppypar}
\end{document}